# Neutron-proton pairing correlations in a single $l-$shell model.*

A. Márquez Romero[a], J. Dobaczewski[a−d], A. Pastore[a]

[a]Department of Physics, University of York, Heslington, York, Y010 5DD, United Kingdom
[b]Department of Physics, P.O. Box 35 (YFL), FI-40014 University of Jyväskylä, Finland
[c]Institute of Theoretical Physics, Faculty of Physics, University of Warsaw, ul. Pasteura 5, PL-02093 Warsaw, Poland
[d]Helsinki Institute of Physics, P.O. Box 64, FI-00014 University of Helsinki, Finland

The long standing problem of neutron-proton pairing correlations is revisited by employing the Hartree-Fock-Bogoliubov formalism with neutron-proton mixing in both the particle-hole and particle-hole channels. We compare numerical calculations performed within this method with an exact pairing model based on the $SO(8)$ algebra. The neutron-proton mixing is included in our calculations by performing rotations in the isospin space using the isocranking technique.

## 1. Introduction

In an analogous way to electrons in superconducting metals, nucleons in nuclei also form Cooper pairs, and thus pairing is a significant feature of nuclear structure [1]. Given the two different fermions, neutrons and protons, that build up the nucleus, three different pairing couplings can be constructed: proton-proton (pp), neutron-neutron (nn), and neutron-proton (np). Typically, only pairing correlations among like-particles are considered.

However, in the region of the nuclear landscape where the numbers of protons and neutrons are similar, the np pairing is expected to play an important role because of the similarity between the proton and the neutron wavefunctions at the Fermi surfaces of both species, cf. [2, 3]. A suitable

---

mean-field description of pairing correlations is given by the Hartree-Fock-Bogoliubov (HFB) method, where the particle-hole and particle-particle channels are treated on the same footing [4].

Because of the affinity of the orbitals that protons and neutrons occupy at the Fermi surface, their wavefunctions overlap and a consistent mean-field theory needs to include the np mixing in both the particle-hole [5, 6] and particle-particle channels. Consequently, single-particle states become those of a nucleon in a superposition of neutron and proton parts. The np mixing is included in our calculations using the cranking model in isospin space (isocranking), through which we can have a complete control over the isospin degree of freedom.

The article is organized as follows: in Section 2 we present the theoretical background of an exact algebraic model for the pairing correlations, HFB formalism, and isocranking technique. In Section 3 we present results and we give our conclusions in Section 4.

## 2. Formalism

In a generalization of nucleon pairing, four different kinds of couplings can be generated: isovector ($T = 1$) pairing, with nn, pp, and np pairs; and isoscalar ($T = 0$) pairing, with np pairs only. A simple and exactly solvable model based upon the $SO(8)$ algebra [7, 8] considers both isovector and isoscalar pairing and can give important insights about the behaviour of both channels. The $SO(8)$ pairing Hamiltonian is here solved numerically using the HFB formalism, which we modify to include the np mixing using the isocranking model.

### 2.1. SO(8) algebraic model

We consider nucleons moving in a single $l-$shell, with spatial degeneracy $2l + 1$. Therefore, the total degeneracy, taking into account the spin and isospin degrees of freedom, equals $\Omega = 4(2l + 1)$. The model Hamiltonian, in the $LST$ coupling scheme, reads

$$H = -g(1-x) \sum_{\nu=-1,0,1} P_\nu^\dagger P_\nu - g(1+x) \sum_{\mu=-1,0,1} D_\mu^\dagger D_\mu, \qquad (1)$$

where pair-creation operators read,

$$P_\nu^\dagger = \sqrt{\frac{2l+1}{2}} \left( a_{l\frac{1}{2}\frac{1}{2}}^\dagger a_{l\frac{1}{2}\frac{1}{2}}^\dagger \right)_{M=0,S_z=0,T_z=\nu}^{L=0,S=0,T=1}, \qquad (2)$$

$$D_\mu^\dagger = \sqrt{\frac{2l+1}{2}} \left( a_{l\frac{1}{2}\frac{1}{2}}^\dagger a_{l\frac{1}{2}\frac{1}{2}}^\dagger \right)_{M=0,S_z=\mu,T_z=0}^{L=0,S=1,T=0}, \qquad (3)$$

and $a^{\dagger}_{l\frac{1}{2}\frac{1}{2}}$ is the creation operator of a particle with angular momentum $l$, spin $\frac{1}{2}$ and isospin $\frac{1}{2}$. $P^{\dagger}$ ($D^{\dagger}$) creates a pair of particles coupled to total angular momentum $L = 0$, total spin $S = 0$ ($S = 1$) and total isospin $T = 1$ ($T = 0$).

The interaction strength is represented by a constant $g$ and $x$ is a mixing parameter that tunes the competition between the isoscalar and isovector channels. By means of the group-theory methods, analytic formulas for the energies can be written for the specific cases of $x = \pm 1, 0$ [7, 8]. The energies as a function of $x$ and for several values of the total spin and isospin of the system can be obtained by means of diagonalizing the Hamiltonian matrix in Eq. (1) (see [9] for details).

## 2.2. HFB formalism with neutron-proton mixing

The HFB formalism [3] relies upon the Bogoliubov transformation from a single-particle basis to a quasiparticle basis defined as

$$\beta^{\dagger}_k = \sum_i v_{ki} a_i + u_{ki} a^{\dagger}_i. \tag{4}$$

For the description of the particle-hole and particle-particle channels we need, in consequence, two densities called the normal density $\rho$ and pairing tensor $\kappa$. The HFB equations are set as a diagonalization problem with

$$\begin{pmatrix} h & \Delta \\ -\Delta^* & -h^* \end{pmatrix} \begin{pmatrix} U \\ V \end{pmatrix} = E_i \begin{pmatrix} U \\ V \end{pmatrix}, \tag{5}$$

where $h = \epsilon + \Gamma$, and

$$\Gamma_{ii'} = \sum_{qq'} \overline{v}_{iq'i'q} \rho_{qq'} \quad \text{and} \quad \Delta_{ii'} = \frac{1}{2} \sum_{qq'} \overline{v}_{ii'qq'} \kappa_{qq'}, \tag{6}$$

are the ph and pp mean fields, respectively, whereas $\overline{v}$ are antisymmetrized matrix elements of the interaction. Matrix $\epsilon$ represents the one-body part of the Hamiltonian, and thus for the pairing Hamiltonian (1) it is equal to zero. Vectors $(U, V)$ contain coefficients $v_{ki}$ and $u_{ki}$ and completely determine transformation (4), and consequently, the HFB wavefunction.

To control average values of the particle number $\hat{N}$ and isospin components, $\hat{T}_x$, $\hat{T}_y$, and $\hat{T}_z$, we need to solve the HFB equations (5) by iterative diagonalization using the Routhian $h'$,

$$h' = h - \lambda \hat{N} - \lambda_x \hat{T}_x - \lambda_y \hat{T}_y - \lambda_z \hat{T}_z, \tag{7}$$

instead of the mean-field $h$, where $\lambda$'s are the set of Lagrange parameters that fix the corresponding expectation values. This constrained minimization is performed using the Augmented Lagrangian Method [10]. Since for vectors $(U, V)$ we do not assume any particular phase convention, we perform all calculations within complex arithmetics.

### 2.3. Cranking in isospin space

The isospin symmetry is controlled in Eq. (7) by means of the Lagrange parameters $\lambda_x$, $\lambda_y$, and $\lambda_z$, which are analogous to those used in the cranking model for the description of rotating nuclei [11], that is, we identify the Lagrange parameters with isocranking frequencies. We use the parametrization $(\lambda_x, \lambda_y, \lambda_z) = (\lambda_0 \sin(\theta), 0, \lambda_0 \cos(\theta))$, whereupon we perform isorotations around axis tilted by angle $\theta$ in the isospace. By changing $\theta$ from 0 to $\pi$, we are able to study the entire multiplet of isobaric analog states [5, 11]. Radius $\lambda_0 = \lambda_z \big|_{\theta=0}$ is determined by adjusting average proton and neutron numbers of the so-called $z$-isoaligned states [5], namely, those with $T_z = \pm T$. Rotations in the isospace are only performed in the $\hat{T}_z$–$\hat{T}_x$ plane, because those involving non-vanishing values of $\hat{T}_y$ are redundant, and would lead to time-reversal symmetry breaking [5].

## 3. Results

For values of $x$ in the range of $[-1, 1]$, we performed ground-state HFB calculations and compared them to: (i) exact solutions of the model and (ii) solutions based on the generalized BCS approach [8]. In Fig. 1, we see that the HFB and BCS formalisms correctly reproduce trends of the exact results, with the HFB results being closer to the exact ones. This is so, because the generalized BCS approach neglects the ph mean field $\Gamma$ (6). For the $\langle \hat{T}_z \rangle = 0$ case shown in Fig. 1, the energy is symmetrical under the change of $x \longrightarrow -x$, which implies a similar behaviour of the isoscalar and isovector channels. For $\langle \hat{T}_z \rangle > 0$, we obtain an increase of energy with increasing $x$, because the isovector pairing correlations then start dominating [8].

When performing isocranking, the energies and average values of the isospin squared do not depend on the isocranking angle $\theta$, because Hamiltonian (1) does not involve any isospin-breaking terms. In Fig. 2, we show results for the first and third components of the isospin, $\langle \hat{T}_x \rangle$ and $\langle \hat{T}_z \rangle$, obtained for the $z$-isoaligned state of $\langle \hat{T}_z \rangle \big|_{\theta=0} = 14$. We see that isovectors $\langle \hat{\vec{T}} \rangle$ and $\langle \hat{\vec{\lambda}} \rangle$ are then parallel to one another, as it is concluded in [5]. The np mixing is effectively included for all values of the isocranking angle $0 < \theta < \pi$. For $\theta = 0$ or $\pi$ we obtain purely neutron or proton states, respectively, and for $\theta = \frac{\pi}{2}$, the neutrons and protons become fully mixed, see an explicit discussion in Ref. [12].

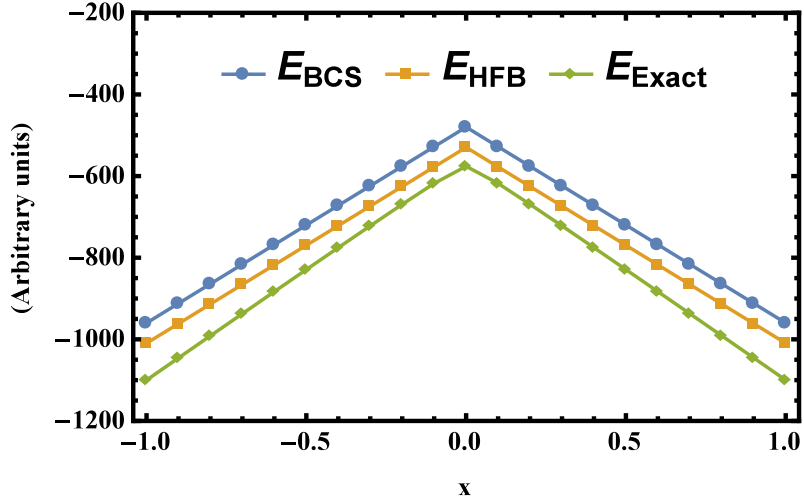

Fig. 1. Total energies (in arbitrary units) as functions of the mixing parameter $x$, obtained from the HFB (squares), generalized BCS (circles), and exact solutions (diamonds) with $l = 15$, $N = 64$, and $\langle \hat{T}_z \rangle = \langle \hat{T}_x \rangle = 0$, corresponding to $\lambda_0 = 0$.

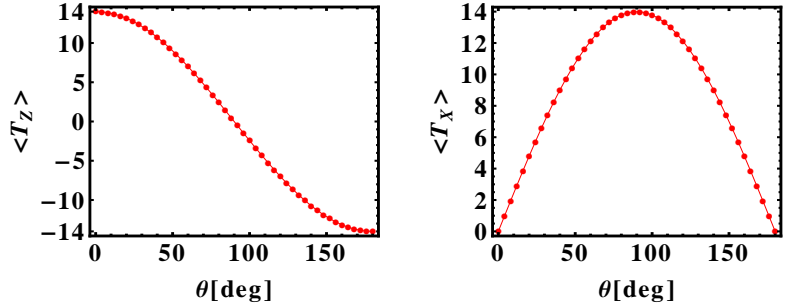

Fig. 2. First and third components of the isospin as functions of $\theta$, obtained for $l = 10$, $N = 36$, and $x = 0$, for an isoaligned state with $\langle \hat{T}_z \rangle \big|_{\theta=0} = 14$.

## 4. Conclusions and perspectives

To treat the neutron-proton pairing correlations in nuclei, we constructed a self-consistent method based on the Hartree-Fock-Bogoliubov formalism including neutron-proton mixing at the mean-field level. We tested this method against results of an exactly solvable model based on the $SO(8)$ algebra, and we confirmed the predicted trends. The neutron-proton mixing was included by means of the cranking in the isospin space. Our approach will become a baseline for studies involving restoration of broken symmetries and/or generator-coordinate mixing of isocranked solutions. In this way, we will analyze beyond-mean-field effects absent in the pure HFB calculations. The final goal will be to port the obtained methodology to the cases of atomic nuclei described within realistic energy density functionals.

## 5. Acknowledgments

This work was supported in part by a Consolidated Grant from the UK Science and Technology Facilities Council (STFC) and by the Academy of Finland and University of Jyväskylä within the FIDIPRO program.

## REFERENCES

[1] A. Bohr, B. R. Mottelson, D. Pines, *Phys. Rev.* **110**, 936 (1958).

[2] Alan L. Goodman, *Phys. Rev. C* **60**, 014311 (1999).

[3] E. Perlińska, S. G. Rohoziński, J. Dobaczewski, W. Nazarewicz, *ibid.* **69**, 014316 (2004).

[4] P. Ring, P. Schuck, *The Nuclear Many-Body Problem*, Springer-Verlag, Berlin 1980.

[5] K. Sato, J. Dobaczewski, T. Nakatsukasa, W. Satuła, *Phys. Rev. C* **88**, 061301(R) (2013).

[6] J. A. Sheikh, N. Hinohara, J. Dobaczewski, T. Nakatsukasa, W. Nazarewicz, K. Sato, *ibid.* **89**, 054317 (2014).

[7] V.K.B. Kota, J.A. Castilho Alcarás, *Nuclear Physics A* **764**, 181 (2006).

[8] J. Engel, S. Pittel, M. Stoitsov, P. Vogel, J. Dukelsky, *Phys. Rev. C* **55**, 1781 (1997).

[9] Sing Chin Pang, *Nuclear Physics A* **128**, 497 (1969).

[10] A. Staszczak, M. Stoitsov, A. Baran, W. Nazarewicz, *Eur. Phys. J. A* **46**, 85 (2010).

[11] Wojciech Satuła, Ramon Wyss, *Phys. Rev. Lett.* **86**, 4488 (2001).

[12] P. Bączyk, J. Dobaczewski, M. Konieczka, W. Satuła, T. Nakatsukasa, K. Sato, *arXiv:1701.04628v3* (2017).